\documentclass[twocolumn,showpacs,preprintnumbers,amsmath,amssymb]{revtex4}
\usepackage{graphicx}% Include figure files
\usepackage{dcolumn}% Align table columns on decimal point
\usepackage{bm}% bold math

\begin{document}

\preprint{APS/123-QED}

\title{Unveiling the intruder deformed 0$^+_2$ state in $^{34}$Si}
\author{F.~Rotaru$^1$, F.~Negoita$^1$, S.~Gr\'evy$^{2*}$,
J.~Mrazek$^3$, S. Lukyanov$^4$, F. Nowacki$^{5}$, A. Poves$^{6}$, O.~Sorlin$^2$, C.~Borcea$^1$, R.~Borcea$^1$,
A.~Buta$^1$, L.~C\'aceres$^2$, S.~Calinescu$^1$, R.~Chevrier$^7$,
Zs. Dombr\'adi$^8$, J.~M.~Daugas$^7$, D.~Lebhertz$^{2}$, Y.~Penionzhkevich$^{4}$,
C.~Petrone$^1$, D.~Sohler$^8$, M.~Stanoiu$^1$ and J.~C.~Thomas$^2$ }

\affiliation{$^1$Horia Hulubei National Institute for Physics and Nuclear Engineering, IFIN-HH, P.O.B. MG-6, 077125 Magurele, Romania,}
\affiliation{$^2$Grand Acc\'el\'erateur National d'Ions Lourds (GANIL), CEA/DSM - CNRS/IN2P3, Bd Henri Becquerel, BP 55027, F-14076 Caen Cedex 5, France}
\affiliation{$^3$Nuclear Physics Institute, AS CR, CZ-25068 Rez, Czech Republic}
\affiliation{$^4$FLNR, JINR, 141980 Dubna, Moscow region, Russia}
\affiliation{$^5$IPHC, Universit\'e de Strasbourg, IN2P3/CNRS; BP28, F-67037 Strasbourg Cedex, France}
\affiliation{$^6$Departamento de F\'isica Te\'orica and IFT-UAM/CSIC, Universidad Aut\'onoma de Madrid, E-28049 Madrid, Spain.}
\affiliation{$^7$CEA, DAM, DIF, Bruy\`eres-le-Ch\^atel, F-91297 Arpajon Cedex, France}
\affiliation{$^8$Institute of Nuclear Research, H-4001Debrecen, Pf.51, Hungary}

\date{\today}

\begin{abstract}
The 0$^+_2$ state in $^{34}$Si has been populated at the {\sc Ganil/Lise3} facility through the $\beta$-decay of a newly discovered 1$^+$ isomer in $^{34}$Al of 26(1)~ms half-life. The simultaneous detection of $e^+e^-$ pairs allowed the determination of the excitation energy E(0$^+_2$)=2719(3)~keV and the  half-life T$_{1/2}$=19.4(7)~ns, from which an electric monopole strength of $\rho^2$(E0)=13.0(0.9)$\times$10$^{-3}$ was deduced.
The 2$^+_1$ state is observed to decay both to the 0$^+_1$ ground state and to the newly observed 0$^+_2$ state (via a 607(2)~keV transition) with a ratio R(2$^+_1$$\rightarrow$0$^+_1$/2$^+_1$$\rightarrow$0$^+_2$)=1380(717).
Gathering all information, a weak mixing with the 0$^+_1$ and a large deformation parameter of $\beta$=0.29(4) are found for the 0$^+_2$ state, in good agreement with shell model calculations using a new {\sc sdpf-u-mix} interaction allowing \textit{np-nh} excitations across the $N=20$ shell gap.

\end{abstract}


\maketitle

\indent
In 1949 Mayer, Haxel, Suess and Jensen~\cite{goe49,hax49} independently gave a
description of the observed shell
gaps at nucleon numbers $2, 8, 20, 28, 50, 82$ and $126$ in terms of mean field potential
including the spin-orbit interaction. With this model, these special numbers -
renamed 'magic numbers'-, as well as the properties of the related nuclei observed
at that time such as spin, magnetic moments, discontinuities in binding energies,
and $\beta$-decay systematics could be explained. Later, other remarkable properties
of magic nuclei have been found: they have a high energy $2^+$ state and a weak
transition probability B(E2:0$^+$$\rightarrow$2$^+$).
The picture of immutable shell gaps persisted until the  ground breaking experiments
performed between 1975 and 1984 in very neutron rich nuclei
close to the neutron magic number $N=20$. Although it was known since long that
the ground state parity of $^{11}$Be was at odds with the naive shell model
picture \cite{Talmi60}, this fact was overlooked until much later.
Studies of charge radii, atomic masses and nuclear spectra in the $_{12}$Mg and
$_{11}$Na isotopic chains have shown that a region of deformation exists at
$N=20$ below $^{34}_{14}$Si~\cite{32Mg_def-exp}. More recently it has been found
that the B(E2) of $^{32}$Mg~\cite{32Mg-coulex} is about 4 times larger than the
one of $^{34}$Si~\cite{Ibbo}, hereby confirming the onset of the regime of
quadrupole collectivity in the region.
In the framework of the shell model, the deformation in $^{32}$Mg was soon
associated with two-particle-two-hole $(2p2h)$ excitations across the $N=20$
shell gap~\cite{32Mg_def-theo}.
These $2p2h$ configurations  were referred to as
intruders since they lie outside the normal model space description of the $sd$
shell nuclei.
The region of those nuclei, the ground state of which is dominantly an intruder
configuration  while their normal configuration ground state is found as an excited
state, is called an "island of inversion". Nuclei around
$^{32}$Mg were proposed first to form such an island of
inversion~\cite{bau89,cau98,hey11}. It has been demonstrated in a recent evaluation
of the experimental data of $^{31}$Mg and $^{33}$Mg~\cite{ney11} that their ground
state wave function is
indeed dominated by two neutrons excitations into the $pf$ orbits. Recent theoretical
works~\cite{Fortune,Hinohara} go a bit further and propose the mixing of the
normal and the intruder states for $^{32}$Mg allowing even for a normal
configuration dominated ground state \cite{Fortune}.
The major pillars to understand the inversion mechanism are the $0^+_{1,2}$
states in $^{30,32}$Mg and $^{34}$Si.
Adding two
neutrons to $^{30}$Mg may provoke the inversion of the normal and intruder
configurations. The latter are expected to be shifted by nearly 3~MeV to
become the ground state of $^{32}$Mg. Along the isotonic chain we anticipate
that the transition is even more abrupt: by removing two protons from
$^{34}$Si, the intruder state has to be shifted down by about
4~MeV with respect to the spherical one to become the ground state of $^{32}$Mg.

Excited $0^+$ states were searched for in $^{30}$Mg, $^{32}$Mg and
$^{34}$Si for a better understanding of the inversion mechanism. Despite many experimental efforts, this quest was vain for about 30 years until
the recent discovery of the 0$^+_2$ states in $^{30}$Mg at
1789~keV~\cite{Schw09} and in $^{32}$Mg at 1058~keV~\cite{wim10}. While the
excited 0$^+$ state in $^{30}$Mg could be assigned to the intruder
configuration~\cite{Schw09}, the assignment of the ground state to the
intruder and the excited 0$^+$ state to the normal configuration in
$^{32}$Mg has been recently questioned~\cite{Fortune2}. Detailed spectroscopy of
$^{34}$Si resulting in the discovery of a $0^+_2$ intruder state is an
important step towards understanding the coexistence of the normal and intruder
configurations~\cite{hey11}. A candidate for the $0^+_2$ state in $^{34}$Si has
been proposed at 2133~keV in~Ref.\cite{num01} but experiments which followed
were not able to confirm this result~\cite{gre05,mit02,iwa03}. In~\cite{iwa03},
a new candidate has been tentatively proposed at 1846~keV, but not confirmed by
later works~\cite{gre05,mit02,Gelin}.

In the present work we propose to use the $\beta$-decay of $^{34}$Al to populate
the 0$^+_2$ state in $^{34}$Si. As $^{34}_{13}$Al$_{21}$ lies at the boundary of
the island of inversion, it should exhibit normal and intruder configurations at
similar excitation energies. Indeed, in the shell model calculations
of~\cite{him08}, its ground state (J$^\pi$=4$^-$) has a mixed configuration $\pi
d_{5/2}\otimes\nu f_{7/2}$ and $\pi d_{5/2}\otimes\nu (d_{3/2})^{-2}(f_{7/2})^3$
while an excited state at $\sim$200~keV (J$^\pi$=$1^+$) has an intruder $2p1h$
configuration $\pi d_{5/2}\otimes\nu (f_{7/2})^2(d_{3/2})^{-1}$
leaving a hole in the neutron $d_{3/2}$ orbit. Following this prediction the
J$^\pi$=$1^+$ state would be a $\beta$-decay isomer. Its decay would mainly
proceed through a Gamow-Teller transition $\nu d_{3/2}\rightarrow\pi d_{5/2}$,
leading mostly to the $2p2h$ 0$^+_2$ state in $^{34}$Si. If the 0$^+_2$ state
is located below the 2$^+_1$ state at 3.326~MeV in $^{34}$Si, it would decay by
an E$0$ transition through internal electron conversion (IC) and/or internal
pair creation (IPF) processes. Thus, electron spectroscopy coupled to
$\beta$-decay spectroscopy was used to search for the 0$^+_2$ state in
$^{34}$Si.

\begin{figure}
\includegraphics[height=8.5cm,angle=0]{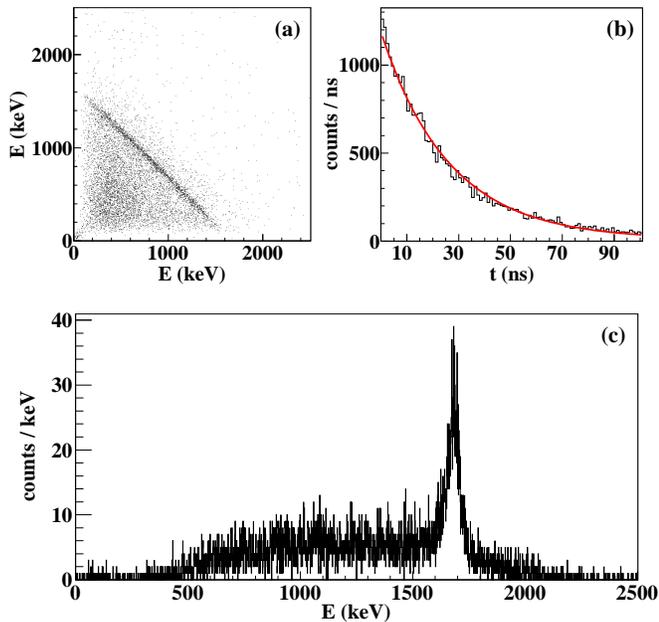}
\caption{\label{fig:telescope} a) Total energy in one telescope (E$_{Si}$+E$_{SiLi}$) versus total energy in another one for events with a
telescope multiplicity $\geqslant$3 and a delay of 16~ns between the $\beta$-trigger and the detected $e^+$ and/or $e^-$. The oblique line
corresponds to a constant energy sum of $e^+e^-$ pairs emitted in the E0 decay of the 0$^+_2$ state in $^{34}$Si.
c) Sum of the energies in both telescopes showing a peak at 1688(2)~keV. A half-life of 19.4(7)ns is deduced for the 0$^+_2$ state from the time difference between a $\beta$-trigger and a $e^+e^-$ pair, as shown in b).}
\end{figure}

The experiment was carried out at the Grand Acc\'el\'erateur National d'Ions Lourds {\sc(Ganil)} facility. The $^{34}$Al nuclei were produced in the fragmentation of a 77.5~A$\cdot$MeV $^{36}$S primary beam of 2~e$\mu$A mean intensity on a 240~mg/cm$^2$ Be target. The {\sc Lise3} spectrometer~\cite{LISE} was used to select and transport the $^{34}$Al nuclei, produced at a rate of 600~pps with a purity of 93\% and a momentum dispersion of 1.48\%.
The produced nuclei were identified on an event by event basis by means of their energy-loss in a stack of Si detectors (labeled Si$_{stack}$) and time-of-flight values referenced to the radio-frequency of the cyclotrons.
The transversal alignment of the $^{34}$Al nuclei was controlled by means of a double-sided Si strip detector located downstream to a 20~degrees-tilted kapton foil of 50~$\mu$m, in which the $^{34}$Al nuclei were implanted. Once the alignment was performed, the implantation depth of the nuclei was adjusted by tilting the Si$_{stack}$ with respect to the beam direction.
Four telescopes (labeled as Si$_{tel}$), each composed of a 1~mm-thick Si detector of 50x50~mm$^2$ followed by a 4.5~mm-thick Si(Li) detector of 45x45~mm$^2$, located 24~mm above and below the beam axis were used to detect electrons and positrons with a geometrical efficiency of $\sim$40\%. In addition two Ge clover detectors of the {\sc Exogam} array, located at 35~mm on the left and right hand sides of the beam axis, were used to detect $\gamma$-rays with an efficiency of 1.6\% at 1~MeV, and 0.8\% at 3.3~MeV.
The experiment ran in sequences of beam-on (120 ms) during which nuclei were collected and beam-off (300 ms) during which the Si$_{tel}$ detected the $\beta$-rays (from the $\beta$-decay) as well as $e^+e^-$ (from IPF). Note that the detection of these particles was also considered in the beam-on mode in anti-coincidence with an ion detected in Si$_{stack}$.
The $0^+_2$ state would decay mainly through IPF if located at a high energy $E_{0^+_2}$ below the $2^+_1$ state at 3.326~MeV. In this hypothesis, the electron and positron would share a total energy E$_{e^-}$+E$_{e^+}$=E$_{0^+_2}-$1022~keV. The search for these events was achieved by requiring a delayed coincidence between three Si$_{tel}$ telescopes. Fig.~\ref{fig:telescope}a shows the total energy in one telescope versus the total energy in another. The oblique line corresponds to events in which the detected energy sum in two telescopes equals to 1688(2)~keV (as shown in Fig.~\ref{fig:telescope}c). Taking into account the energy losses of the $e^+e^-$ pair in the kapton foil as well as their energy-angle correlations~\cite{hof90} with {\sc Geant4} simulations~\cite{gea4}, we deduce that the total energy of the emitted pair (E$_{e^-}$+E$_{e^+}$) was 9(1)~keV higher, establishing a 0$^+_2$ state at E$_{0^+_2}$=2719(3)~keV in $^{34}$Si.

As shown in the Fig.~\ref{fig:telescope}b, a half-life of $T_{1/2}(E0)$=19.4$\pm$0.7~ns has been obtained for the $0^+_2$ state from the time difference between a $\beta$-ray in one of the Si$_{tel}$ and a pair detected in another. A consistent value of  19.2$\pm$0.8~ns was found from the time difference between a $\beta$-ray and a $\gamma$-ray of 511~keV, arising  from the annihilation of the positron at rest, detected in {\sc Exogam}.
Therefore, the transition electric monopole strength  $\rho^2$(E0:0$^+_2$$\rightarrow$0$^+_1$)=13.0(0.9)$\times$10$^{-3}$ is calculated using the
internal conversion $\Omega_{IC}$=1.331x10$^7$~s$^{-1}$ and internal pair creation $\Omega_{IPF}$=2.733x10$^9$~s$^{-1}$ transition rates. These values have been obtained from the one of ref.~\cite{a} extrapolated to A=34 and corrected to take into account the atomic screening. The detailed procedure can be found in~\cite{force_thesis}.

\begin{figure}[t]
\includegraphics[height=8.5cm,angle=270]{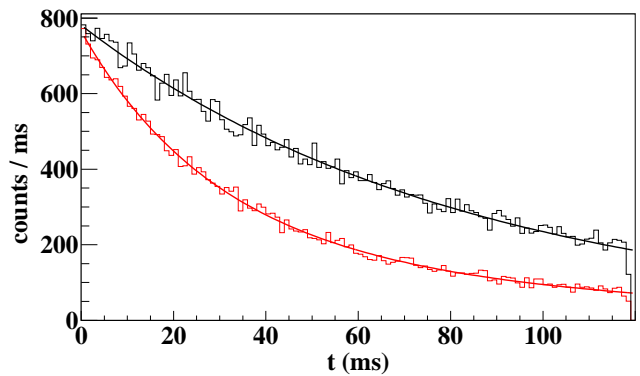}
\caption{\label{fig:t12} $\beta$-decay time spectra obtained in coincidence with the 926~keV (in black) and 511~keV (in red) $\gamma$-rays of $^{34}$Si giving different half-lives corresponding to the 4$^-$ ground state [54.4(5)5~ms] and the 1$^+$ isomeric state [26(1)~ms] in $^{34}$Al.}
\end{figure}

The existence of two $\beta$-decaying states in $^{34}$Al is proven in the
present work by the fact that half-lives obtained when $\beta$'s are followed by
926~keV or 511~keV $\gamma$-rays differ significantly as shown in
Fig.~\ref{fig:t12}. The transition at 926~keV is due to the
4$^-$$\rightarrow$3$^-$ $\gamma$-decay, as shown in~\cite{num01}. Its half-life
of 54.4(5)~ms agrees well with the value of 56.3(5)~ms obtained in~\cite{num01}.
Conversely, the transition of 511~keV corresponding to the
1$^+$$\rightarrow$0$^+_2$ $\beta$-decay has a significantly shorter half-life of
26(1)~ms. The half-lives of the 4$^-$ and 1$^+$ states in $^{34}$Al compare well
with the values of 59 and 30~ms predicted by shell model calculations, a direct feeding of the $0^+_2$ state of 17\% being predicted
from the J$^\pi$=1$^+$.
As no $\gamma$-ray (except a large number of 511~keV due to positrons annihilation) was observed in coincidence with the $\sim$2$\times$10$^4$ $e^+e^-$ events selected in Fig.~\ref{fig:telescope} (a,c) we surmise that the $0^+_2$ state is fed \emph{directly} by the $\beta$-decay of the 1$^+$ isomer of $^{34}$Al. However, an absolute direct decay-branch to the $0^+_2$ state is hard to obtain as the ratio of isomeric feeding in $^{34}$Al could not be determined. As for the direct feeding of the $2^+_1$ state in $^{34}$Si through the decay of the J=1$^+$ isomer, the situation is more complex since all states populated in the decay of the $4^-$ state- transit through it. Since the $\beta$-decay lifetime in coincidence with the $2^+_1\rightarrow0^+_1$ transition (49.8(2)~ms) is shorter than the one obtained in~\cite{num01}, it is concluded that the 2$^+_1$ state is also fed (directly and/or indirectly) from $\beta$-decay of the isomer in $^{34}$Al. A J$^\pi$=1$^+$ value is assigned to the $\beta$-decaying isomer in $^{34}$Al by virtue of comparison to shell model calculations and $\beta$-decay selection rules.

\begin{figure}[t]
\includegraphics[width=8.5cm,angle=0]{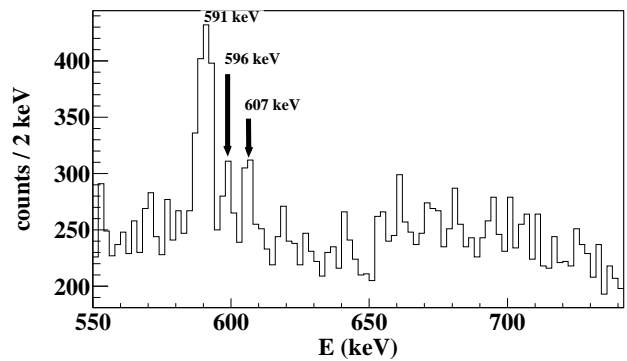}
\caption{\label{fig:gamma} Part of the gamma energy spectrum following the implantation of $^{34}$Si nuclei. The main peak corresponds to the known 591~keV transition in $^{34}$Si. Peaks at 607~keV and 596~keV  correspond to the 2$^+_1$$\rightarrow$0$^+_2$ decay and the (n,n'$\gamma$) reactions on the $^{74}$Ge nuclei of {\sc Exogam} detectors, respectively.}
\end{figure}

Energy wise the $2^+_1$ state in $^{34}$Si could decay both to the $0^+_2$ state (located 607~keV below) and to the $0^+_1$ ground state leading to the known 3.326~MeV transition.
Observation of both decay branches inform on the degree of mixing between these states. Shell model predict B(E2:2$^+_1\rightarrow$0$^+_2$)=67~e$^2$fm$^4$ and B(E2:2$^+_1\rightarrow$0$^+_1$)=11~e$^2$fm$^4$. When weighted by the E$_\gamma ^5$ factor for E2 transitions, the expected branching to the $0^+_2$ state represents only 0.12\% of the total decay of the $2^+_1$ state. To observe the weak decay branch through the 607~keV transition it was necessary to reduce the $\gamma$-background. This was achieved by requiring a multiplicity M$_{Si_{tel}}\ge2$.
In Fig.~\ref{fig:gamma}, the 607~KeV transition is seen together with the known 591~keV $\gamma$-line from the 4970~keV state in $^{34}$Si~\cite{num01}. When the beta-decay of $^{34}$Al occurs to unbound states in $^{34}$Si, the emitted neutrons can react with the $^{74}$Ge nuclei contained in the {\sc Exogam} detectors and excite its $2^+_1$ state at 595.8~keV, giving rise to an enlarged peak at the corresponding energy in Fig.~\ref{fig:gamma}.

Despite a weak signal to noise ratio obtained for the 607~keV peak, a ratio R(2$^+_1$$\rightarrow$0$^+_1$/2$^+_1$$\rightarrow$0$^+_2$)=1380(717) has been extracted for the decay of the 2$^+_1$ state to the 0$^+_2$ and 0$^+_1$ states taking into account the $\gamma$ efficiencies at 607~keV and 3.326~MeV and the Si detector efficiencies with the related uncertainties. A value of B(E2:2$^+_1$$\rightarrow$0$^+_2$)=61(40)~e$^2$fm$^4$ is deduced using the measured value of  B(E2:2$^+_1$$\rightarrow$0$^+_1$)=17(7)~e$^2$fm$^4$~\cite{ibb98} determined via Coulomb excitation.

Information on the mixing and deformation of the 0$^+_{1,2}$ states in $^{34}$Si can be obtained using a two level mixing model assuming spherical $\beta_S$ and deformed $\beta_D$ configurations, as it has been done for example in~\cite{for10}. Using the relation B(E2:2$^+_1$$\rightarrow$0$^+_1$)/B(E2:2$^+_1$$\rightarrow$0$^+_2$)$\sim$tan$^2\theta$~\cite{mac89}, a weak mixing ratio of cos$^2\theta$=0.78(9) is deduced from the experimental B(E2) values. We remind here that the maximum mixing ratio would lead to cos$^2\theta$=0.5. The magnitude of the electric monopole matrix element can be written as a function of the mixing ratio and the difference of shapes, $\beta_S$ and $\beta_D$, of the two configurations before mixing~\cite{Wood99}, $\rho^2$(E0)=(3Ze/4$\pi$)$^2$sin$^2\theta$cos$^2\theta$($\beta_D^2$-$\beta_S^2$)$^2$. Using the experimental value of the mixing ratio, the experimental electric monopole strength is reproduced when deformation parameters of $\beta_D$=0.29(4) and $\beta_S$=0 are taken.
\begin{table}[tb]
\caption{Comparison between the experimental and shell model energies (in keV) and reduced transition probabilities (in e$^2$fm$^4$) for  $^{34}$Si, $^{32}$Mg and $^{30}$Mg.}
\label{table:3}
\begin{tabular}
{@{}l|cc|cc|cc}

                                 & $^{34}$Si &       & $^{32}$Mg  &       & $^{30}$Mg\\
                                 & exp.      & s.m.  & exp.       & s.m.  & exp.      & s.m.\\
\hline
E(0$^+_2$)                       & 2719(3)   & 2570  & 1058(2)$^a$& 1282  & 1788.2(4) & 1717 \\
E(2$^+_1$)                       &3326(1)$^b$& 3510  & 885.3(1)   & 993   & 1482.8(3) & 1642 \\
B(E2:2$^+_1$$\rightarrow$0$^+_1$)& 17(7)$^c$ & 11    & 91(16)$^d$ & 85    & 48(6)$^e$  & 59 \\
B(E2:2$^+_1$$\rightarrow$0$^+_2$)& 61(40)    & 67    & $\le$109$^a$& 15    & 11(1)$^f$ & 9 \\
 \hline
\end{tabular}\\[2pt]
${^a}$:~\cite{wim10}; ${^b}$:~\cite{iwa03,num01}; ${^c}$:~\cite{ibb98}; ${^d}$:~\cite{32Mg-coulex};
${^e}$:~\cite{Nied05}; ${^f}$:~\cite{Schw09}
\end{table}

\indent We compare now  the experimental results with the shell model calculations performed with the code {\sc Antoine} \cite{rmp05} using the effective interaction {\sc sdpf-u-mix} which is an extension of  {\sc sdpf-u-si} \cite{now09}.
The {\sc sdpf-u-si} interaction was designed for 0$\hbar\omega$ calculations of very neutron rich $sd$ nuclei around $N=28$ in a valence space comprising the full $sd$($pf$)-shell for the protons(neutrons), {\it i.e.} this interaction was defined with a core of $^{28}$O. Its single particle energies (SPE's) and monopoles (neutron-proton $sd$-$pf$ and neutron-neutron $pf$-$pf$) were fixed by the spectra of $^{35}$Si, $^{41}$Ca, $^{47}$K and $^{49}$Ca. In order to allow for the mixing among different \textit{np-nh} neutron configurations across $N=20$, it is necessary to add to {\sc sdpf-u-si} the following new ingredients: a) The off-diagonal cross shell $sd$-$pf$ matrix elements, which are taken from the Lee-Kahana-Scott G-matrix \cite{LKS} scaled as in ref.~\cite{ca40}; b) The neutron SPE's on a core of $^{28}$O: for the the $sd$-shell orbits we use always the USD values \cite{USD}, while for the $pf$-shell orbits we have no experimental guidance at all. Nonetheless, for any particular set  of  $pf$-shell SPE's, the neutron-neutron  $sd$-$pf$ monopoles must be chosen such as to reproduce the spectrum of $^{35}$Si and the $N=20$ gap. We have anchored our choice to the energy of the first excited 0$^+$ state in $^{30}$Mg, because this guarantees that in our isotopic course toward $N=20$ the descent of the intruder states proceeds with the correct slope. Indeed, at  0$\hbar\omega$ {\sc sdpf-u-mix} and {\sc sdpf-u-si} produce identical results.

The results of the calculations performed with this new {\sc sdpf-u-mix} interaction are gathered in Table~\ref{table:3}. There is a very nice agreement for the excitation energies and B(E2)'s in $^{34}$Si using the standard $sd$-shell effective charges $e_\pi$=1.35$e$ and $e_\nu$=0.35$e$. The 0$^+_1$ ground state has 89\% of neutron closed shell configuration whereas the excited 0$^+_2$ and 2$^+_1$ are built on $2p2h$ excitations at 86\%. Thus the image of coexistence between a closed-shell 0$^+_1$ and  a strongly correlated 0$^+_2$ state stands for $^{34}$Si.
It is worthwhile to mention that, as illustrated in Table~\ref{table:3}, the results obtained with this new interaction for the $^{30,32}$Mg agree also very well with the experimental data. The ground state of $^{30}$Mg is built on normal configurations at 77\% and its first 0$^+$ excited state is an intruder with the same proportion (77\%). The situation is more complex in $^{32}$Mg, with the ground state being dominated by intruder configurations at 88\%  whereas the first excited 0$^+$ is an even mixture of normal and intruder components. With these mixing ratios, the dramatic shift observed for the intruder configuration in $^{32}$Mg with respect to both $^{30}$Mg and $^{34}$Si is well reproduced.

Concerning  $^{34}$Al, the calculation produces the right ground state spin 4$^-$, a first excited 5$^-$ at 0.25~MeV and a 1$^+$ isomer at 0.55~MeV. The lifetimes of the ground state (59 ms) and the isomer (30 ms) agree nicely with the experimental data (54.4(5) and 26(1), respectively).
The multiplet of negative parity states is dominated by the neutron $1p0h$ configuration $(f_{7/2})^{+1}$ with a proton hole in $d_{5/2}$ consistent with the doubly magic picture of $^{34}$Si. The mixing in the 4$^-$ ground state, discussed in~\cite{him08}, is calculated to be around 22\%. The structure of the isomeric 1$^+$ state is, as expected, dominated (92\%) by the neutron $2p1h$ configuration $(d_{3/2})^{-1}(f_{7/2})^2$.

\indent To summarize, the $\beta$-decay of a newly discovered 1$^+$ isomer in $^{34}$Al (T$_{1/2}$=26(1)~ms) has been used to populate and study for the first time the 0$^+_2$ state at 2719(3) keV in $^{34}$Si. From the spectroscopic information $-$ $\rho^2$(E0:0$^+_2$$\rightarrow$0$^+_1$)=13.0(0.9)$\times$10$^{-3}$ and B(E2:2$^+_1$$\rightarrow$0$^+_2$)=61(40) e$^2$fm$^4$ $-$ a weak mixing ratio of 0.78(9) with the 0$^+_1$ state and a large deformation parameter $\beta$=0.29(4) are extracted. Therefore the spherical ground state 0$^+_1$ and the deformed 0$^+_2$ state coexist in $^{34}$Si.
State of the art shell model calculations using the new {\sc sdpf-u-mix} interaction accounting for the mixing of normal states with \textit{np-nh} excitations across the $N=20$ shell gap has been performed, the results of which are in very good agreement with the experimental data. These calculations show a $12-22$\% mixing of the intruder component to the normal one in the ground states of $^{30}$Mg and  $^{34}$Si, respectively, while a similar admixture of the normal configurations to the intruder ones is calculated in the ground state of $^{32}$Mg as well as in the $0_2^+$ states of $^{30}$Mg and $^{34}$Si. Thus the basic idea of the island is confirmed in the framework of the shell model, although the picture became a more refined via allowing for configuration mixing.\\

\indent The authors are thankful to
the {\sc Lise} staff for their effort in the preparation of the
experiment. This work has been supported by BMBF 06BN109, GA
of Czech Republic 202/040791, MICINN-FPA2009-13377,
CM-HEPHACOS-S2009/ESP-1473 (Spain),
the EC through the Eurons project contract RII3-CT-3/2004-506065,
OTKA K68801 and K100835, the Bolyai Foundation and by Romanian ANCS, CNCS
UEFISCDI, project number PN-II-RU-TE-2011-3-0051.
RB, AB, FN and FR acknowledge the IN2P3/CNRS support.\\

\noindent $^*$corresponding author: grevy@in2p3.fr. Present affiliation : Centre d'Etudes Nucl\'eaire de Bordeaux Gradignan {\sc (cenbg)}, UMR 5797 CNRS/IN2P3, Chemin du Solarium, BP 120, 33175 Gradignan Cedex, France\\

\end{document}